\documentclass[english]{ceurart}
\usepackage[utf8]{inputenc}
\usepackage[T1]{fontenc}
\usepackage{babel}
\usepackage{stmaryrd}

\usepackage{bussproofs}
\EnableBpAbbreviations

\usepackage{graphbox}
\DeclareRobustCommand{\DLLogo}{%
  \begingroup\normalfont
  \kern-1.75pt\includegraphics[align=c,height=1.25\baselineskip]{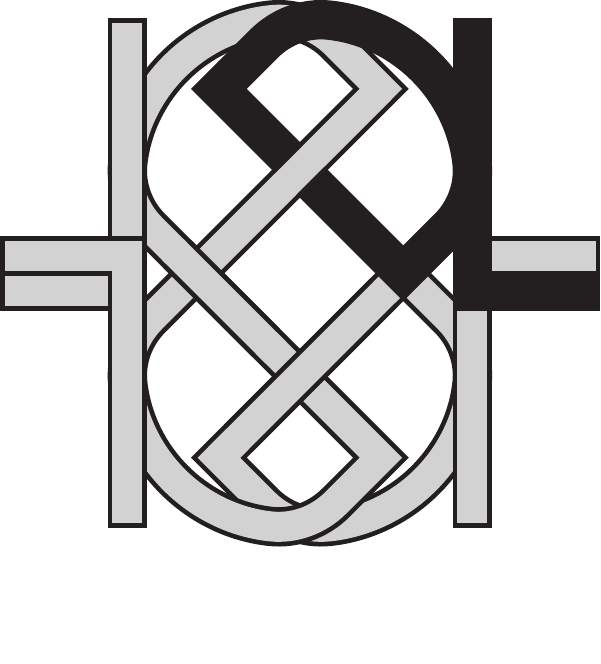}\kern-1.5pt%
  \endgroup
}

\usepackage{amsthm}

\usepackage{tikz}
\usetikzlibrary{shapes,snakes}

\usepackage{amssymb}
\usepackage{amsmath}
\usepackage{xspace}
\usepackage{todonotes}
\usepackage{hyperref}
\usepackage{mathtools}
\usepackage{tikz}
\usepackage{graphicx}
\usepackage{csquotes}
\usepackage{pgfplots}
\usepackage{framed}
\usepackage{subcaption}

\newcommand{\ttodo}[4]{\ifthenelse{\equal{#1}{inline}}{\todo[inline,author=#2,color=#3]{#4}}{\todo[color=#3]{#2: #4}}}

\def\define#1#2#3%
{%
\renewcommand*{\do}[1]{%
 \expandafter\newcommand\csname
 #1\endcsname{#2}
}
\docsvlist{#3}
}

\define{#1}
{{\textsc{#1}}\xspace}
{Elk,Lethe,Fame,Evonne,Evee,HermiT}
\newcommand{\LIBRARYNAME}{\Evee-\textsc{libs}\xspace}
\newcommand{\PLUGINNAME}{\Evee-\textsc{protege}\xspace}

\newcommand{\wrt}{w.r.t.\ }
\newcommand{\ie}{i.e.,\xspace}
\newcommand{\eg}{e.g.\ }

\newcommand{\vs}{vs.\ }

\define{#1mc}
{{\ensuremath{\mathcal{#1}}}\xspace}
{A,B,C,D,E,F,G,H,I,J,K,L,M,N,O,P,Q,R,S,T,U,V,W,X,Y,Z}

\define{#1}
{{\textsc{#1}}\xspace}
{NP,coNP,PSpace,ExpTime,ExpSpace}

\define{#1}
{{\ensuremath{\mathcal{#1}}}\xspace}
{ALC,EL,ALCOI,ALCH,ELH,ELI,SROIQ}

\newcommand{\ELp}{\ensuremath{\mathcal{EL}^+}\xspace}

\newcommand{\Heur}{\texttt{HEUR}\xspace}

\newcommand{\Extr}{\texttt{EXTR}\xspace}

\newcommand{\pluginTwoScreenshots}{
	\begin{figure}
	\centering
	\begin{subfigure}{.5\textwidth}
		\includegraphics[width=0.95\linewidth]{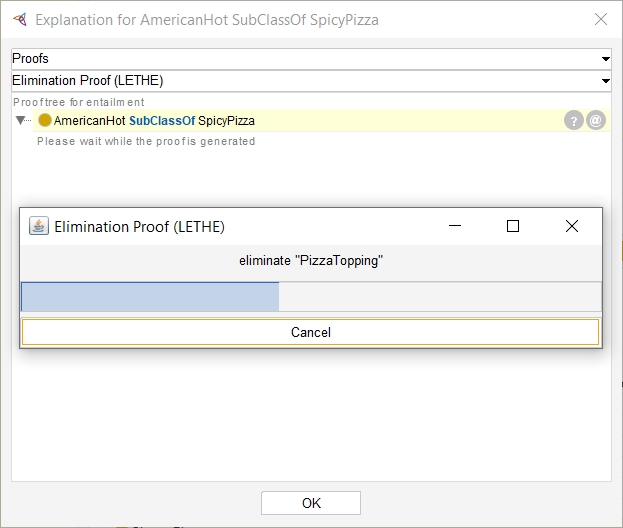}
		\centering
	\end{subfigure}%
	\begin{subfigure}{.5\textwidth}
		\includegraphics[width=0.95\linewidth]{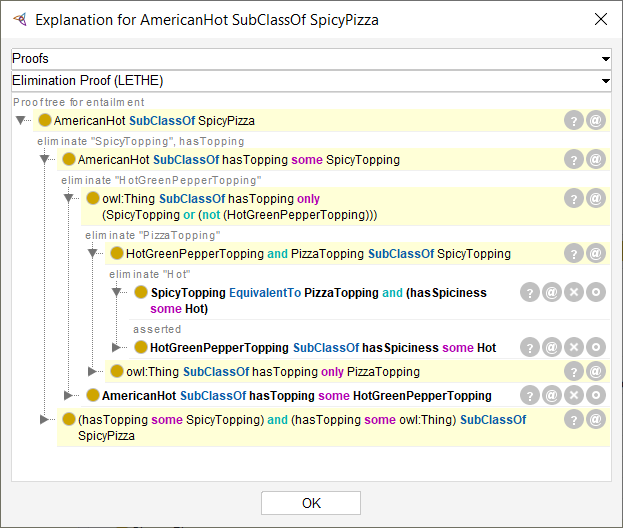}
		\centering
	\end{subfigure}
		\caption{Loading screen during proof generation (left) and a finished proof (right)}
		\label{fig:Plugin-Screenshot}
	\end{figure}
}

\newcommand{\plugin}{plug-in\xspace}
\newcommand{\Plugin}{Plug-in\xspace}
\newcommand{\plugins}{plug-ins\xspace}

\newcommand{\Protege}{Prot{\'e}g{\'e}\xspace}
\tikzset{marks/.style={only marks, mark size = 1pt, solid, opacity=.7}}

\newcommand{\diagramFontSize}{\small}

\newcommand{\plotFBlog}[6]{
    \diagramFontSize
    \begin{tikzpicture}
        \begin{axis}[height=0.43\textwidth,
            width=0.43\textwidth,
            title={#1},
            xlabel={#2},
            ylabel={#3},
            xmode=log,
            ymode=log,
            every axis title/.append style={
                at={(0.5,.95)}
            },
            every axis y label/.style={
                at={(ticklabel cs:0.5)},rotate=90, anchor=near ticklabel,
            },
            every axis x label/.style={
                at={(ticklabel cs:0.5)}, anchor=near ticklabel,
            },
            xmin=99,
            xmax=#4,
            ymin=99,
            ymax=#4,
            grid=major,
            domain = 100:#4,
            xtick scale label code/.code={},
            ytick scale label code/.code={},
            xtick distance=10,
            ytick distance=10,
            ]
            \addplot[draw=black,samples at={100,1000000}] {x};
            \addplot[
                marks,
                green!60!black!60!,
                scatter,
                visualization depends on={value \thisrowno{2} \as \count},
                scatter/@pre marker code/.code={
                \def\markopts{
                    mark size=1+\count/10
                }
                \expandafter\scope\expandafter[\markopts]
                },
                scatter/@post marker code/.code={
                    \endscope
                }
            ] table [x index = 0, y index = 1, col sep = comma] {#5};
            \addplot[
                marks,
                mark = triangle*,
                red!60!black!60!,
                scatter,
                visualization depends on={value \thisrowno{2} \as \count},
                scatter/@pre marker code/.code={
                \def\markopts{
                    mark size=1+\count/10
                }
                \expandafter\scope\expandafter[\markopts]
                },
                scatter/@post marker code/.code={
                    \endscope
                }
            ] table [x index = 0, y index = 1, col sep = comma]  {#6};
        \end{axis}
    \end{tikzpicture}
}

\newcommand{\plotFB}[9]{
    \diagramFontSize
    \begin{tikzpicture}
        \begin{axis}[height=0.43\textwidth,
            width=0.43\textwidth,
            title={#1},
            xlabel={#2},
            ylabel={#3},
            every axis title/.append style={
                at={(0.5,.95)}
            },
            every axis y label/.style={
                at={(ticklabel cs:0.5)},rotate=90, anchor=near ticklabel,
            },
            every axis x label/.style={
                at={(ticklabel cs:0.5)}, anchor=near ticklabel,
            },
            xmin=0,
            xmax=#4,
            ymin=0,
            ymax=#4,
            domain = 0:#4,
            grid=major,
            xtick scale label code/.code={},
            ytick scale label code/.code={},
            xtick distance=10,
            ytick distance=10,
            legend style={font=\small},
            legend pos=#9,
            legend cell align={left},
            ]
            \addplot[draw=black] {x};
            \addplot[
                marks,
                green!60!black!60!,
                scatter,
                visualization depends on={value \thisrowno{2} \as \count},
                scatter/@pre marker code/.code={
                \def\markopts{
                    mark size=1+\count/10
                }
                \expandafter\scope\expandafter[\markopts]
                },
                scatter/@post marker code/.code={
                    \endscope
                }
            ] table [x index = 0, y index = 1, col sep = comma] {#5};
            \addplot[
                marks,
                mark = triangle*,
                red!60!black!60!,
                scatter,
                visualization depends on={value \thisrowno{2} \as \count},
                scatter/@pre marker code/.code={
                \def\markopts{
                    mark size=1+\count/10
                }
                \expandafter\scope\expandafter[\markopts]
                },
                scatter/@post marker code/.code={
                    \endscope
                }
            ] table [x index = 0, y index = 1, col sep = comma]  {#6};
            \legend{,#7,#8}
        \end{axis}
    \end{tikzpicture}
}

\newcommand{\plotLElog}[5]{
    \diagramFontSize
    \begin{tikzpicture}
        \begin{axis}[height=0.43\textwidth,
            width=0.43\textwidth,
            title={#1},
            xlabel={#2},
            ylabel={#3},
            xmode=log,
            ymode=log,
            every axis title/.append style={
                at={(0.5,.95)}
            },
            every axis y label/.style={
                at={(ticklabel cs:0.5)},rotate=90, anchor=near ticklabel,
            },
            every axis x label/.style={
                at={(ticklabel cs:0.5)}, anchor=near ticklabel,
            },
            xmin=99,
            xmax=#4,
            ymin=99,
            ymax=#4,
            grid=major,
            domain = 100:#4,
            xtick scale label code/.code={},
            ytick scale label code/.code={},
            xtick distance=10,
            ytick distance=10,
            ]
            \addplot[draw=black,samples at={100,1000000}] {x};
            \addplot[
                marks,
                green!60!black!60!,
                scatter,
                visualization depends on={value \thisrowno{2} \as \count},
                scatter/@pre marker code/.code={
                \def\markopts{
                    mark size=1+\count/10
                }
                \expandafter\scope\expandafter[\markopts]
                },
                scatter/@post marker code/.code={
                    \endscope
                }
            ] table [x index = 0, y index = 1, col sep = comma] {#5};
        \end{axis}
    \end{tikzpicture}
}

\newcommand{\plotLE}[5]{
    \diagramFontSize
    \begin{tikzpicture}
        \begin{axis}[height=0.43\textwidth,
            width=0.43\textwidth,
            title={#1},
            xlabel={#2},
            ylabel={#3},
            every axis title/.append style={
                at={(0.5,.95)}
            },
            every axis y label/.style={
                at={(ticklabel cs:0.5)},rotate=90, anchor=near ticklabel,
            },
            every axis x label/.style={
                at={(ticklabel cs:0.5)}, anchor=near ticklabel,
            },
            xmin=0,
            xmax=#4,
            ymin=0,
            ymax=#4,
            domain = 0:#4,
            grid=major,
            xtick scale label code/.code={},
            ytick scale label code/.code={},
            xtick distance=10,
            ytick distance=10,
            ]
            \addplot[draw=black] {x};
            \addplot[
                marks,
                green!60!black!60!,
                scatter,
                visualization depends on={value \thisrowno{2} \as \count},
                scatter/@pre marker code/.code={
                \def\markopts{
                    mark size=1+\count/10
                }
                \expandafter\scope\expandafter[\markopts]
                },
                scatter/@post marker code/.code={
                    \endscope
                }
            ] table [x index = 0, y index = 1, col sep = comma] {#5};
        \end{axis}
    \end{tikzpicture}
}

\newif\ifextendedVersion

\extendedVersiontrue %

\begin{document}

\copyrightyear{2022}
\copyrightclause{Copyright for this paper by its authors.
  Use permitted under Creative Commons License Attribution 4.0
  International (CC BY 4.0).}

\conference{\DLLogo{} DL 2022: 35th International Workshop on Description Logics,
  August 7--10, 2022, Haifa, Israel}

\ifextendedVersion
\title{On~the~Eve~of~True~Explainability~for~OWL~Ontologies: Description Logic Proofs with 
Evee and Evonne (Extended Version)}
\else
\title{On~the~Eve~of~True~Explainability~for~OWL~Ontologies: Description Logic Proofs with 
Evee and Evonne}
\fi
\author[1]{Christian Alrabbaa}[%
orcid=0000-0002-2925-1765,
email=christian.alrabbaa@tu-dresden.de,
]
\author[1]{Stefan Borgwardt}[%
orcid=0000-0003-0924-8478,
email=stefan.borgwardt@tu-dresden.de,
]
\author[1]{Tom Friese}[%
email=tom.friese@tu-dresden.de,
]
\author[1]{Patrick Koopmann}[%
orcid=0000-0001-5999-2583,
email=patrick.koopmann@tu-dresden.de,
]
\author[2]{Juli\'an M\'endez}[%
orcid=0000-0003-1029-7656,
email=julian.mendez2@tu-dresden.de,
]
\author[1]{Alexej Popovič}[%
email=alexej.popovic@tu-dresden.de,
]
\address[1]{Institute of Theoretical Computer Science, Technische Universität Dresden, 01062 
Dresden, Germany}
\address[2]{Interactive Media Lab Dresden, Technische Universität Dresden, 01062 Dresden, 
Germany}

\begin{abstract}
  When working with description logic ontologies, understanding
entailments derived by a description logic reasoner is not
always straightforward.
  So far, the standard ontology editor \Protege offers two services
to help: (black-box) justifications for OWL 2 DL ontologies, and (glass-box)
proofs for lightweight OWL EL ontologies, %
where the latter exploits the proof facilities of reasoner \Elk. Since justifications are often 
insufficient in explaining inferences, there is thus only little tool support for explaining
inferences in more expressive DLs.
  In this paper, we introduce \LIBRARYNAME, a Java library for computing
proofs for DLs up to \ALCH, and \PLUGINNAME, a collection of
\Protege plugins for displaying those proofs in \Protege. We also give a
short glimpse of the latest version of \Evonne, a more advanced standalone
application for displaying and interacting with proofs computed with
\LIBRARYNAME.

\end{abstract}


\maketitle

\section{Introduction}

Description logics (DLs)~\cite{BHLS-17} have gained popularity through the standardization of the Web Ontology Language OWL\footnote{\url{https://www.w3.org/TR/owl2-overview/}} and the development of an OWL Java API~\cite{DBLP:journals/semweb/HorridgeB11}, editing tools, and reasoners.
This popularity comes with the need of explaining description logic reasoning to domain experts who are not familiar with DLs.
We consider here the problem of explaining the entailment of
logical consequences from DL
ontologies (as opposed to explaining \emph{non}-consequences, which requires other 
techniques~\cite{DBLP:conf/ki/AlrabbaaHT21,DBLP:conf/kr/HaifaniKT21}).
Research on explaining description logic inferences in the beginning considered \emph{proofs} that provide detailed inference steps through which the consequence can be obtained~\cite{DeMc-96,DBLP:conf/ecai/BorgidaFH00}.
In many cases, especially in light-weight DLs, it can be sufficient to compute a \emph{justification}, a minimal set of ontology axioms that entail the consequence~\cite{DBLP:conf/ijcai/SchlobachC03,DBLP:conf/ki/BaaderPS07,Horr-11}.
However, if justifications become very large or the ontology is formulated in a more expressive 
DL, providing intermediate inference steps between a justification and its consequence may be 
required for understanding.
Generating proofs is achievable via heuristic search for possible intermediate 
inferences~\cite{DBLP:conf/semweb/HorridgePS10}, concept 
interpolation~\cite{DBLP:conf/jelia/Schlobach04}, 
forgetting~\cite{DBLP:conf/lpar/AlrabbaaBBKK20}, or directly using the inference rules 
underlying a consequence-based reasoner like \Elk~\cite{DBLP:journals/jar/KazakovKS14}, 
which however only supports \ELp.
Recently, a \plugin for the ontology editor 
\Protege~\cite{DBLP:journals/aimatters/Musen15} %
 was developed that shows proofs generated by \Elk in a human-readable 
form~\cite{DBLP:conf/dlog/KazakovKS17}.
Related work investigated the complexity of finding proofs that are optimal \wrt various measures, \eg proof size or depth~\cite{DBLP:conf/lpar/AlrabbaaBBKK20,DBLP:conf/dlog/AlrabbaaBBKK20,DBLP:conf/cade/AlrabbaaBBKK21}.
For example, from a set of instantiated inference rules, \eg computed by \Elk, one can extract a proof of minimal depth in polynomial time~\cite{DBLP:conf/cade/AlrabbaaBBKK21}.

In this paper, we present a collection of tools for proof generation, also for expressive DLs other than \ELp.
Our Java library \LIBRARYNAME (EVincing Expressive Entailments)
implements various proof generation methods that form the basis for \PLUGINNAME, 
a collection of \Protege \plugins{} for interactively inspecting proofs, and
\Evonne, a more advanced proof visualization tool.
This library extends our previously reported implementations of proof generation 
algorithms~\cite{DBLP:conf/lpar/AlrabbaaBBKK20,IJCAR22} in various ways.
One can extract proofs that are minimized \wrt size or depth from the output of
\Elk, generate \emph{elimination
proofs}~\cite{DBLP:conf/lpar/AlrabbaaBBKK20,IJCAR22} that can additionally be
optimized \wrt several parameters, or generate more \emph{detailed proofs}
directly via the resolution-based calculus used in the uniform
interpolation~(UI) tool \Lethe~\cite{DBLP:journals/ki/Koopmann20}.
Elimination proofs can support different DLs, depending on the UI tool that is
used internally in a black-box fashion --- our library currently uses a version
of \Fame~\cite{DBLP:conf/cade/ZhaoS18} that supports \ALC, and \Lethe, which
supports \ALCH. While we discussed and evaluated the different
methods for elimination proofs before~\cite{IJCAR22}, the method for detailed
proofs is new. We provide a quantitative comparison of this method with the
elimination proofs based on ontologies from BioPortal.

To use the proofs in practice to support ontology engineers with explanation
services, we offer two tools to show proofs generated by \LIBRARYNAME to users.
\PLUGINNAME
utilizes PULi (the Proof Utility Library) and the \Protege
proofs \plugin~\cite{DBLP:conf/dlog/KazakovKS17} to show the proofs directly
in the ontology editor \Protege. This might be the easiest way for users to use
our services, and allowed us to perform a qualitative user study for our
different proof methods presented in this paper.
\Evonne is a stand-alone web
application, early prototypes of which were presented
in~\cite{DBLP:conf/dlog/AlrabbaaBDFK20,DBLP:conf/semweb/FlemischLAD20}. Since
then, it has seen many visual improvements and new features, and offers now many
different ways of displaying and
interacting with proofs, together with a visualization of the modular structure
of ontologies, and guidance for repairing undesired entailments. A larger
publication
discussing and evaluating the design decisions made in \Evonne is currently
under preparation. We here just give a brief overview of its proof visualization
facilities.
The source code and installation instructions for \LIBRARYNAME and \PLUGINNAME
can be found at
\url{https://github.com/de-tu-dresden-inf-lat/evee}, where also the plugins can
be downloaded. \Evonne can be tested online and downloaded
under~\url{https://imld.de/evonne/}.

\section{Preliminaries}

We consider a consequence $\Omc\models\alpha$ that is to be explained, where \Omc is a DL ontology~\cite{BHLS-17} and $\alpha$ an axiom.
A \emph{justification} for $\alpha$ in $\Omc$ is a subset-minimal $\Jmc\subseteq\Omc$ such that $\Jmc\models\alpha$~\cite{DBLP:conf/ijcai/SchlobachC03}.
In the worst case, there can be  exponentially many justifications for an entailment.
Following~\cite{DBLP:conf/lpar/AlrabbaaBBKK20}, we define \emph{proofs} of 
$\Omc\models\alpha$ as finite, acyclic, directed hypergraphs, where vertices~$v$ are labeled 
with axioms~$\ell(v)$ and hyperedges are of the form $(S,d)$, with $S$ a tuple of vertices and 
$d$ a vertex such that $\{\ell(v)\mid v\in S\}\models\ell(d)$; the leafs of a proof must be 
labeled by elements of~\Omc and the root by~$\alpha$.
We depict hyperedges either using horizontal bars (\eg $\frac{~p~~~~p\to q~}{q}$) or using graphs with two kinds of nodes (see \eg Figure~\ref{fig:elimination-proof}).
In this paper, all proofs are \emph{trees}, \ie no vertex can appear in the first component of multiple hyperedges.
The \emph{size} of such a proof is the number of its vertices (this is called \emph{tree size} in~\cite{DBLP:conf/lpar/AlrabbaaBBKK20,DBLP:conf/cade/AlrabbaaBBKK21}), and its \emph{depth} is the length of the longest path from a leaf to the root.
Given a finite collection of valid inference steps (hyperedges), one can extract a (tree) proof that is composed from a subset of these steps and has minimal size or minimal depth in polynomial time~\cite{DBLP:conf/lpar/AlrabbaaBBKK20,DBLP:conf/cade/AlrabbaaBBKK21}.

\section{Proof Generation with \LIBRARYNAME}\label{Sec:proof-types}

\LIBRARYNAME implements several (families of) proof generation methods with
different advantages. They all use the Java interface
\texttt{IProofGenerator<OWLAxiom>} with a method \texttt{getProof} that takes as
input an \texttt{OWLAxiom} and returns an \texttt{IProof<OWLAxiom>} that can
either be serialized into JSON format using a \texttt{JsonProofWriter}, explored
using the method \texttt{getInferences}, or measured using an
\texttt{IProofEvaluator}.
For entailments that only depend on axioms in~\ELp, we implemented a
method that optimizes proofs generated by \Elk using various
proof measures (in the \textit{evee-elk-proof-extractor} library). For \ALC and \ALCH, we implemented two other proof generation
methods, \emph{elimination proofs} (\textit{evee-elimination-proofs}) and \emph{\Lethe-based proofs} (\textit{evee-lethe-proof-extractor}).

\subsection{%
Optimized \Elk Proofs}

We implemented the Dijkstra-like proof search algorithm
from~\cite{DBLP:conf/lpar/AlrabbaaBBKK20,DBLP:conf/dlog/AlrabbaaBBKK20,DBLP:conf/cade/AlrabbaaBBKK21} that takes as input a set of instantiated inference rules
that prove a target consequence and outputs an optimized proof.
This works for any \emph{recursive} measure of proof quality---intuitively, measures that can be computed recursively over the tree structure of the proof.
Examples include the (tree) size
weighted size, and depth of a proof.
We apply this algorithm to the output of \Elk, providing a different view on \Elk proofs than 
the existing \plugin~\cite{DBLP:conf/dlog/KazakovKS17}, which may show multiple ways to prove a DL axiom at the same time.
Due to the fast reasoning with \Elk, and because our algorithm runs only on
the relatively small input extracted from \Elk, this is the fasted proof
generation method in \LIBRARYNAME.
An experimental comparison of size-minimal \Elk proofs and elimination proofs can be found in~\cite{DBLP:conf/lpar/AlrabbaaBBKK20}.

\subsection{Elimination Proofs}

For entailments that require more expressive DLs, such as \ALC, we implemented
the forgetting-based approach from~\cite{DBLP:conf/lpar/AlrabbaaBBKK20}. We now call
those proofs \emph{elimination proofs}, because they perform inferences
by eliminating names. Inferences in an elimination proof look as follows:
\begin{center}
  \AXC{$\alpha_1$}
  \AXC{$\dots$}
  \AXC{$\alpha_n$}
  \RL{eliminate $X$ }
  \TIC{$\beta$}
  \DP
\end{center}
where $X$ is a concept or a role name, $\{\alpha_1$, $\ldots$,
$\alpha_n\}\models\beta$, no subset of
$\{\alpha_1$, $\ldots$, $\alpha_n\}$ entails~$\beta$, and $X$ does not occur in~$\beta$. The general idea is that $\beta$ is derived from the
premises $\alpha_1$, $\ldots$, $\alpha_n$ by performing inferences on $X$.
Since usually the
axiom we want to prove contains less names than the axioms it is derived from,
this allows for a step-wise sequence of inferences from the ontology to the 
entailment.
The result is a more high-level explanation than the other proof types.
Elimination proofs do not depend on a specific logic --
provided we have a method to compute elimination steps.

We compute elimination proofs by making use of existing tools for
\emph{forgetting}: %
given
an ontology $\Omc$ and a name $X$, the result of forgetting $X$ from $\Omc$ is
an ontology $\Omc^{-X}$ entailed by~$\Omc$ that %
does not contain~$X$, and preserves all entailments of $\Omc$ that can be expressed without
using~$X$. Forgetting has bad worst-case 
complexities~\cite{DBLP:conf/ijcai/LutzW11,DBLP:journals/ai/NikitinaR14}, and
may not always be successful, for instance if $\Omc$ contains cycles. Nonetheless,
forgetting-tools like \Fame and \Lethe often perform well in practice when
applied to realistic 
ontologies~\cite{DBLP:journals/ki/Koopmann20,DBLP:conf/cade/ZhaoS18}.
Since forgetting is not always possible, these tools may introduce fresh names
to simulate cyclic dependencies in the result.
\Fame may even not return a result at all because it is not complete.

We implemented three methods for computing elimination proofs using forgetting.
All three start
from a justification~\Jmc\footnote{This is computed using the black-box justification algorithm~\cite{DBLP:conf/sum/HorridgePS09} with \HermiT~\cite{DBLP:journals/jar/GlimmHMSW14}.} for the entailment $\Omc\models\alpha$ to be explained, and then
generate a sequence of ontologies $\Jmc=\Omc_1$, $\ldots$ $\Omc_n$ by eliminating
the names that do not occur in~$\alpha$ one after the other. For
robustness, we internally use a wrapper for the chosen forgetting method --- if
the forgetting method \enquote{fails} to eliminate~$X$ (exceeds a fixed timeout set to 10 
seconds by default, 
contains
a definer, or fails for any other reason), we keep~$X$ and continue forgetting the next name. The
wrapper also makes sure that forgetting results are cached.
From this sequence of ontologies, the elimination proof
is then constructed using justifications: specifically, for~$i>1$, we construct
an inference step for $\alpha\in\Omc_i\setminus\Omc_{i-1}$ by taking an
arbitrary\footnote{We also implemented the possibility of looking for the best
justification in each case. However, this adversely affected the running time
and did not lead to significantly smaller proofs.} justification for $\alpha$
from $\Omc_{i-1}$. To further optimize the presentation, we added a technique
to merge several inference steps into one: in an inference
$\frac{~\alpha_1~~\ldots~~\alpha_n~}{\beta}$, we may replace premises $\alpha_i$
with the premises used in the inference producing $\alpha_i$, provided that the
number of premises
in the resulting inference step does not increase. This often leads to easier
proofs in which
several names are eliminated at the same time, \eg using the following
inference:
\begin{center}
  \AXC{$A\sqsubseteq \forall r.(B_1\sqcap B_2\sqcap B_3)$}
  \AXC{$\exists r.(B_1\sqcap B_2\sqcap B_3)\sqsubseteq B$}
  \RL{eliminate $B_1$, $B_2$, $B_3$ }
  \BIC{$A\sqcap\exists r.\top\sqsubseteq B$}
  \DP
\end{center}

\begin{figure}
 \centering
    \begin{tikzpicture}

        \node[draw,rounded corners=5pt,very thick] (c1c3c2) at (0,0)
{$C_1\sqsubseteq C_3\sqcup \mathbf{C_2}$};

        \node[draw,rounded corners=5pt,very thick] (c2c3) at (5,0)
{$\mathbf{C_2}\sqsubseteq C_3$};

        \node[draw] (ec2) at (2.5,-0.25)  {elim. $\mathbf{C_2}$};

        \node[draw,rounded corners=5pt] (c1c3) at (2.5,-1)
{$\mathbf{C_1}\sqsubseteq C_3$};

        \draw (c1c3c2) -> (ec2);
        \draw (c2c3) -> (ec2);
        \draw (ec2) -> (c1c3);

        \node[draw,rounded corners=5pt,very thick] (aarc1) at (-2.5,-1)
{$A\sqsubseteq \forall r.\mathbf{C_1}$};

        \node[draw] (ec1) at (0,-1.25) {elim. $\mathbf{C_1}$};
        \node[draw,rounded corners=5pt] (aarc3) at (0,-2)
        {$A\sqsubseteq \forall \mathbf{r}.\mathbf{C_3}$};

        \draw (aarc1) -> (ec1);
        \draw (c1c3) -> (ec1);
        \draw (ec1) -> (aarc3);

        \node[draw, rounded corners=5pt,very thick] (arc3b) at (5,-2)
{$\forall \mathbf{r}.\mathbf{C_3}\sqsubseteq B$};
        \node[draw] (ec3r) at (2.5, -2.25) {elim. $\mathbf{r}$, $\mathbf{C_3}$};
        \node[draw, rounded corners=5pt,very thick] (ab) at (2.5,-3)
{$A\sqsubseteq B$};

        \draw (aarc3) -> (ec3r);
        \draw (arc3b) -> (ec3r);
        \draw (ec3r) -> (ab);
    \end{tikzpicture}
    \caption{Elimination proof. Premises and conclusion highlighted, eliminated
symbols in bold face.}
    \label{fig:elimination-proof}
\end{figure}
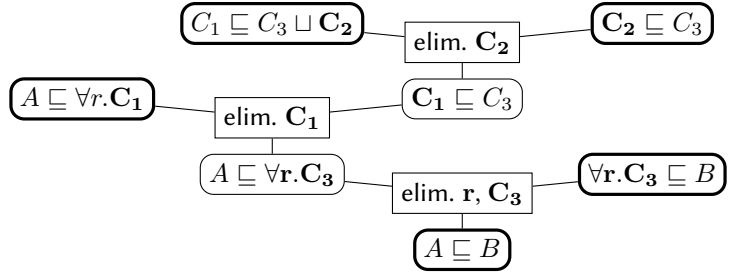

For generating the sequence of ontologies, we employed three strategies. 1) The 
\emph{heuristic method} selects names using the heuristic
described in~\cite{DBLP:conf/lpar/AlrabbaaBBKK20}, 2) the \emph{name-optimized
method} uses
best-first search to optimize the order in which names are eliminated with the aim 
of finding a shortest sequence of ontologies (thus, essentially minimizing the number 
of names being eliminated), 3) the \emph{size-optimized method} also
uses best-first search,
but optimizes the final proof rather based on other measures than the length of
the sequence: size and weighted size.
Forgetting role names early often leads to very short sequences and small proofs, but also 
hides inferences. For this reason, we added the
additional constraint that role
names can only be eliminated if they 
only occur in role restrictions without other names 
(\eg $\exists r.\top$, $\forall r.\bot$). An example of an elimination proof
generated using the heuristic approach is shown in
Figure~\ref{fig:elimination-proof}.
\Lethe and \Fame may reformulate axioms that do not involve the forgotten name
due normalization, which results in elimination inferences that do not really
eliminate a symbol.
For example, such a inference could produce
$C\sqsubseteq D$
from $C\equiv D$. Note that the inference pattern above allows this. Because
they improve readability of the proof, we keep those inferences, but call them
\enquote{normalization} rather than elimination inferences.

We note that the version of \Fame we use only supports \ALC-TBoxes, while the
version of \Lethe supports \ALCH-TBoxes. The generated proofs often look
quite different because \Lethe puts more emphasis on \enquote{beautifying}
the resulting axioms, that is, reformulating them to look more human-readable. Thus,
elimination proofs generated using \Lethe may contain axioms that are easier to
read than those generated using \Fame. On the other hand, it can happen as a result 
of reformulating axioms that the relation between the premises and the conclusion of 
an inference step becomes less obvious.

Details and an experimental evaluation of the optimized algorithms can be found
in~\cite{IJCAR22}.

\subsection{Detailed Proofs Extracted from \Lethe}

Elimination proofs provide high-level explanations, since the detailed
inferences for eliminating a symbol are hidden.
For situations where this
representation is too coarse, we provide a method for
generating more detailed proofs, which uses the forgetting tool \Lethe in a
different way.
When forgetting names in \ALCH TBoxes, \Lethe internally uses a consequence-based 
reasoning procedure
based on a normalized representation of axioms and the calculus shown in
Figure~\ref{fig:lethe-calculus}~\cite{DBLP:conf/lpar/KoopmannS13}. To explain
$\Omc\models A\sqsubseteq B$, we forget all names other than $A$ and $B$ from
$\Omc$, and track the inferences performed by \Lethe. The final set of clauses
will contain a clause from which $A\sqsubseteq B$ can be straight-forwardly
deduced.

\begin{figure}
    \newcommand{\quant}{{\mathsf{Q}}}
    \newcommand{\gr}[1]{{\color{gray!50}#1}}

    \begin{framed}
        \begin{flushleft}
        \begin{minipage}{0.38\textwidth}
            \AXC{$\gr{\top\sqsubseteq}\ C_1\sqcup A$}
            \AXC{$\gr{\top\sqsubseteq}\ C_2\sqcup \neg A$}
            \RL{}
            \BIC{$\gr{\top\sqsubseteq}\ C_1\sqcup C_2$}
            \DP
        \end{minipage}%
        \begin{minipage}{0.315\textwidth}
            \AXC{$\gr{\top\sqsubseteq}\ C\sqcup \exists r.D$}
            \AXC{$r\sqsubseteq s$}
            \RL{}
            \BIC{$\gr{\top\sqsubseteq}\ C\sqcup \exists s.D$}
            \DP
        \end{minipage}%
        \begin{minipage}{0.315\textwidth}
            \AXC{$\gr{\top\sqsubseteq}\ C\sqcup \forall r.D$}
            \AXC{$s\sqsubseteq r$}
            \RL{}
            \BIC{$\gr{\top\sqsubseteq}\ C\sqcup \forall s.D$}
            \DP
        \end{minipage}
        
        \bigskip

            \AXC{$\gr{\top\sqsubseteq}\ C_1\sqcup\exists r.D_1$}
            \AXC{$\gr{\top\sqsubseteq}\ C_2\sqcup\forall s.D_2$}
            \RL{\ where $\Omc\models r\sqsubseteq s$}
            \BIC{$\gr{\top\sqsubseteq}\ C_1\sqcup C_2\sqcup\exists r.(D_1\sqcap D_2)$}
            \DP

        \bigskip

            \AXC{$\gr{\top\sqsubseteq}\ C_1\sqcup\forall r_1.D_1$}
            \AXC{$\gr{\top\sqsubseteq}\ C_2\sqcup\forall r_2.D_2$}
            \RL{\ where $\Omc\models s\sqsubseteq r_1, s\sqsubseteq r_2$}
            \BIC{$\gr{\top\sqsubseteq}\  C_1\sqcup C_2\sqcup\forall s.(D_1\sqcap D_2)$}
            \DP

        \bigskip

        \begin{minipage}{0.7\textwidth}
            \AXC{$\gr{\top\sqsubseteq}\ C_1\sqcup\exists r_1.D_1$\quad
            $\gr{\top\sqsubseteq}\ C_2\sqcup\forall r_2.D_2$\quad
            $\ldots$\quad
            $\gr{\top\sqsubseteq}\ C_n\sqcup\forall r_n.D_n$}
            \RL{}
            \UIC{$\gr{\top\sqsubseteq}\ C_1\sqcup \ldots\sqcup C_n$}
            \DP
        \end{minipage}%
        \begin{minipage}{0.3\textwidth}
            \raggedright
            where $\Omc\models \bigsqcap_{i=1}^{n}D_i\sqsubseteq\bot$,
            $\Omc\models r_1\sqsubseteq r_2, \ldots, r_1\sqsubseteq r_n$
        \end{minipage}%
        \vspace*{-\baselineskip}%
    \end{flushleft}%
    \end{framed}%
    \caption{Inference rules of \Lethe for \ALCH TBoxes. The last rule 
    uses \HermiT for satisfiability testing.}
    \label{fig:lethe-calculus}
\end{figure}

\begin{figure}
    \centering
       \begin{tikzpicture}

        
        \node[draw,rounded corners=5pt,very thick] (c1c32) at (3.5,0)
        {$C_1\sqsubseteq C_3\sqcup C_2$};

        
        \node[draw,rounded corners=5pt,very thick] (c2c3) at (6,0)
{$C_2\sqsubseteq C_3$};


        \node[draw,rounded corners=5pt] (c1c3) at (4.75,-0.75)
        {$C_1\sqsubseteq C_3$};
        \draw (c1c32) -> (c1c3);
        \draw (c2c3) -> (c1c3);


        \node[draw,rounded corners=5pt] (c3c3) at (7.25,-0.75)
        {$C_3\sqcap \neg C_3\sqsubseteq \bot$};


        \node[draw,rounded corners=5pt] (c3c1) at (6,-1.5) {$
C_1\sqcap\neg C_3 \sqsubseteq\bot\phantom{)}$};
        \draw (c1c3) -> (c3c1);
        \draw (c3c3) -> (c3c1);
        

        \node[draw,rounded corners=5pt,very thick] (ac3b) at (1,0)
        {$\forall r.C_3\sqsubseteq B$};


        \node[draw,rounded corners=5pt] (berc3) at (1,-0.75)
        {$\top\sqsubseteq B\sqcup\exists r.\neg C_3$};

        \draw (ac3b) -> (berc3);


        \node[draw,rounded corners=5pt,very thick] (aarc1) at (-2,-0.75)
        {$A\sqsubseteq\forall r.C_1$};

        
        \node[draw,rounded corners=5pt] (aberc3c1) at (-0.25,-1.5)
        {$A\sqsubseteq B\sqcup\exists r.(C_1\sqcap \neg C_3)$};

        \draw (aarc1) -> (aberc3c1);
        \draw (berc3) -> (aberc3c1);


        \node[draw,rounded corners=5pt,very thick] (ab) at (3, -2.25)
        {$A\sqsubseteq B$};

        \draw (aberc3c1) -> (ab);
        \draw (c3c1) -> (ab);

       \end{tikzpicture}
       \caption{Detailed proof generated based on inference logging in \Lethe
(rules omitted).}
       \label{fig:detailed-proof}
   \end{figure}
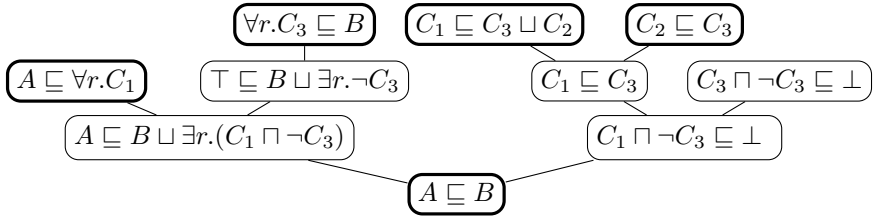

We modified the code of \Lethe so that it
logs all performed inferences, including information on the side conditions. 
In addition to applying the inferences 
from Figure~\ref{fig:lethe-calculus}, normally \Lethe repeatedly normalizes and denormalizes relevant parts of the ontology, and 
applies further simplifications to keep the current representation of the 
forgetting result small. For the logging to operate properly, we added a logging 
mode where all optimizations are turned off, and the entire ontology is normalized 
once at the beginning.  This leads to slower reasoning performance, which is
okay since we apply the method only on justifications rather than large
ontologies.
To avoid complicated inferences using the last rule in Figure~\ref{fig:lethe-calculus},
we make sure that inferences on role names are performed at the very end. At this 
stage, all relevant information for the unsatisfiability is usually already 
propagated into a single existential role restriction, which means that the 
inference has only one premise and the side conditions are not needed anymore. 
Indeed, we found that the resulting inference steps were usually very
straightforward.

To produce a proof from the logging information, some additional steps are
necessary: 1) the normalization as well as the reasoning procedure introduce
fresh names, which have to be replaced again by the concepts they represent, 2)
in the normal form, axioms are always represented as concept disjunctions,
which is not very user-friendly, and 3) the normalized axioms still have to be
linked to (non-normalized) input axioms, as well as to the conclusion.
For 3), we again use justifications computed using \HermiT. For 2), we use the
\enquote{beautification} functionality of \Lethe %
 to produce
more human-friendly forgetting results. This involves
applying some obvious simplifications of concepts ($A\sqcup\neg
A\equiv A\sqcup\top\equiv\forall
r.\top\equiv\top$, $A\sqcap\neg A\equiv A\sqcap\bot\equiv \exists r.\bot\equiv
\bot$), pushing negations inside complex concepts, and moving disjuncts to the left-hand
side of a GCI if this allows to eliminate further negation symbols
(eg. $A\sqsubseteq B\sqcup\exists r.\neg C$ $\Rightarrow$ $A\sqcap\forall
r.C\sqsubseteq B$). Applying those simplifications too rigorously, however, may
again hide inferences performed by \Lethe. We thus \enquote{beautify} the
axioms \emph{before} we replace the introduced names for~1), so that inferences
performed on introduced names are still easily visible. The effect of this can
be seen in the proof in Figure~\ref{fig:detailed-proof}, where negations that
would otherwise be \enquote{beautified away} are still shown.
After transforming the
logged inferences in this way, we use our Dijkstra-style proof search algorithm
to extract a single proof.

\subsubsection{Comparing Elimination Proofs to \Lethe-based Proofs}

\begin{figure}
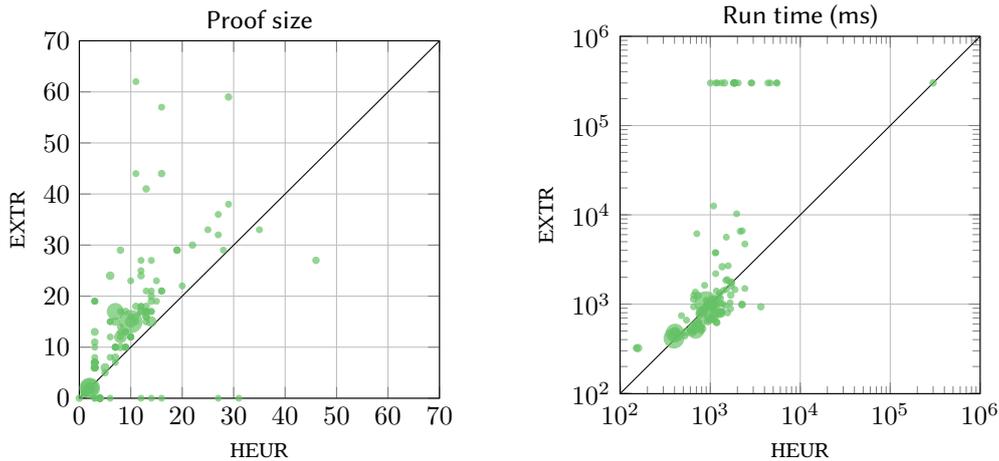

  \centering
  \plotLE{Proof size\vphantom{()}}{\Heur}{\Extr}{70}%
  {figures/LETHE-heur-LETHE-extr-treesize-adj2.csv}%
  \hfil
  \plotLElog{Run time (ms)}{\Heur}{\Extr}{1000000}%
  {figures/LETHE-heur-LETHE-extr-totaltime-adj.csv}%
  \caption{Run time and proof size for the heuristic elimination approach (using \Lethe) \vs detailed proof extraction from \Lethe. Marker size indicates how often each pattern occurred in the BioPortal snapshot.}
  \label{fig:lethe-eval}
\end{figure}

We compared the elimination proofs constructed using \Lethe and the heuristic method (\Heur) with the detailed proofs extracted directly from \Lethe (\Extr) on a collection of proof tasks from the 2017 snapshot of BioPortal~\cite{BIOPORTAL} (see also~\cite{IJCAR22}).
Up to isomorphism, we extracted 138 distinct~\ALCH proof tasks (consisting of an entailed axiom 
and the union of all its justifications) that are not fully expressible in \ELH, representing 327 
different entailments within the BioPortal ontologies.
We then computed proofs using the two methods, imposing a timeout of 5 minutes. This was successful for 325/327 entailments using \Heur.
Due to the deactivated optimizations, \Extr timed out in 24/327 cases, which
mostly happened when the signature of the task was large (more than
\textasciitilde$18$ symbols).
Users should thus use \Extr only if the signature of the justification is
sufficiently small.
In Figure~\ref{fig:lethe-eval}, we compare the proof size (left) and the run time (right) of both methods.
On average, the proof size of \Extr was 90\% higher than for \Heur, but the
average inference step complexity\footnote{This is measured using the
justification complexity from~\cite{DBLP:journals/kbs/HorridgeBPS13}.} was 10\%
lower, supporting our expectation that \Extr computes more detailed proofs.
Interestingly, the \emph{maximum} step complexity of \Extr was 29\% higher,
indicating that the elimination proofs can hide complex inferences.

\section{\Evee \Protege Plugins}

\pluginTwoScreenshots

The easiest way to use our proof generators is via \plugins for \Protege, a popular editor for OWL ontologies~\cite{DBLP:journals/aimatters/Musen15}.
In \Protege, users can
navigate the concept and role hierarchies of the
ontology. When selecting a concept or role name,
further information is shown, such as annotations and asserted or inferred equivalent concepts and superconcepts.
To explain reasoning
results in \Protege, one can click on the \enquote{?}-button next to an
entailment. The standard explanation consists of a list of justifications for the axiom, but this has been extended to support proofs via the
\textit{protege-proof-explanation}\footnote{\url{https://github.com/liveontologies/protege-proof-explanation}}
\plugin, which %
relies on the \textit{proof utility 
library}\footnote{\url{https://github.com/liveontologies/puli}} (PULi)~\cite{DBLP:conf/dlog/KazakovKS17}.

Our proof generators are available in several \Protege \plugins under the umbrella name \PLUGINNAME, utilizing this existing functionality.
The \plugins can be installed by copying their JAR-files into \Protege's \texttt{plugin} folder, where the elimination proof methods are available in two versions for \Lethe and \Fame.
Each proof generation method is registered as a
\emph{ProofService} for the \textit{protege-proof-explanation} \plugin, which can be accessed from a drop-down list in the explanation dialog of \Protege (see Figure~\ref{fig:Plugin-Screenshot}).
Once selected, a ProofService starts generating the actual proof and a progress
bar shows updates to the user.
After proof generation is completed, the result can be explored interactively in a tree view.
By clicking on the grey triangle next to an axiom, the proof steps that lead to this conclusion can be recursively unraveled.

Since proof generation can %
take some time to complete, users
can abort the process by clicking %
 the \enquote{Cancel}-button or
by closing the progress window.
The proof generators that are based on search will then
assemble the best proof %
found so far, provided they already found one.
If this is the case, a warning message appears to inform the user that the
proof may be sub-optimal.

To facilitate modularity, the common functionalities of the proof services are
implemented in \emph{evee-protege-core}, which %
is not a
\plugin itself (in the sense that it does not provide any extension to \Protege).
Rather, evee-protege-core exports a collection of Java class files to
\Protege that are required by the actual \plugins.
Each of our \plugins that provides a ProofService %
uses these classes by inheriting 
from \emph{AbstractEveeProofService} and instantiating one of our proof~generators.

In order to add a ProofService for \Protege, the method 
\emph{getProof} needs to be implemented.
This method returns a \emph{DynamicProof} which in turn needs to implement a method 
\emph{getInferences} to provide the actual inferences to \Protege.
Our \plugins define this method in the class \emph{EveeDynamicProofAdapter} of 
evee-protege-core.
The dynamic nature of the proof comes~in very handy as it allows for the proof generation to be 
executed in a background thread, and to later update the proof explanation window with the 
constructed proof.
To our knowledge~\cite{DBLP:conf/dlog/KazakovKS17}, a dynamic proof was
originally intended to be automatically updated after an asserted axiom is
deleted.
For this, a DynamicProof utilizes \emph{ChangeListeners} which can be informed
about any changes made to a proof that is currently shown to the user.
Our \plugins use this feature by first showing a proof consisting only of a 
single \enquote{placeholder} inference that is displayed when the proof generation is started.
This inference has as its conclusion the axiom for which the proof is created and as its name a 
short message indicating that the proof is currently being computed (see 
Figure~\ref{fig:Plugin-Screenshot}).
Once the proof generation is complete, the method \emph{inferencesChanged} is called on all registered ChangeListeners, which causes \Protege to update the UI with the actual proof.
If no proof can be shown due to a cancellation or if an error occurred during proof generation, the 
\enquote{placeholder} inference will be updated accordingly.

\subsection{User Study}
We conducted a small qualitative user study to evaluate the usability of the \plugins, and to 
validate the variety of proof generation options that \PLUGINNAME provides.
We held individual interviews online, where participants installed the \plugins, 
performed five tasks,\footnote{Tasks and ontologies used in the study: 
\url{https://cloud.perspicuous-computing.science/s/cy5Y3kj94AGYSDB}} and answered 
questions related to these tasks. 
Each task specified an entailment, for 
which the participants were asked to generate two proofs using different methods, and 
compare them in terms of ease of
\ifextendedVersion
understanding. In the appendix, we depict all proofs used in the study.
\else
understanding (See the extended version~\cite{extendedVersion} for all proofs used in the 
study).
\fi
Lastly, participants could provide feedback about what additional features they would like to 
have. In total, we interviewed $10$ participants, all of which often work with Description Logics.

In Tasks~1 and~2, participants were asked to compare proofs generated for an
atomic concept inclusion
in the Skin
Phsyiology Ontology
(SPO).\footnote{\url{https://bioportal.bioontology.org/ontologies/SPO}}
For Task~1, they compared the detailed \Lethe-based proof with the elimination
proof generated using \Lethe with the heuristic method.
Asked which proof is the easier to understand,
$60\%$ chose the elimination proof, %
$10\%$ %
the detailed proof %
and $30\%$ found no difference between the two.
Most of the participants that found no difference stated that it took them the same amount of 
effort to understand both proofs. For the other participants, %
the structure of proofs and axioms and the type of expressions used were the
reasons to prefer one proof over the other.

For Task~2, participants compared elimination proofs generated by \Lethe
and \Fame using the heuristic method.
Here, $50\%$ prefered the \Fame proof and $30\%$ the \Lethe proof in terms of
understandability,
while $20\%$ of the participants found no differences
between them. %
Participants that prefered the \Fame proof pointed out a contradiction caused
by the interplay between a universal and an existential restriction.
For these participants, inferences like that are easier to follow. At the same
time, the lack of such interplay in the \Lethe proof, which used only
existential restrictions, is the reason why most of the other participants
prefered this proof. %

In Task~3, participants were asked to prove a concept unsatisfiability %
in a modified version of the pizza
ontology\footnote{\url{https://protege.stanford.edu/ontologies/pizza/pizza.owl}
} using the proofs provided by the original \Elk proof
plugin~\cite{DBLP:conf/dlog/KazakovKS17} and the size-optimized elimination
proof using \Fame.
Asked which \emph{inference steps} were easier to understand,
$70\%$ chose the \Elk proof,
$20\%$ preferred the elimination proof, and
$10\%$ found no difference. 
However, when asked if they agree that a large proof with simple inferences is easier to 
understand than a smaller proof with intricate inferences, the answers varied. 
Some agreed with the statement, some agreed only as long as a certain ratio between the size of 
a proof and the difficulty of its inferences is not exceeded. Some pointed out that despite their 
agreement they think that people will eventually get used to the more complex inferences and 
how they work. On the other hand, there were some participants who totally disagreed with the 
statement. The main reason is that a short proof provides a better overview.

For the last two tasks, participants looked into proofs for an atomic concept
inclusion in
the BioTop
ontology.\footnote{\url{https://bioportal.bioontology.org/ontologies/BT}}
In Task~4, the comparison was
between the
original Elk plugin and the \Elk proofs optimized for size. In Task~5,
the comparison was between the optimized \Elk Proof and
detailed proof generated by \Lethe. For both tasks, the answers were unanimous.
Participants found
the easier proofs to be the optimized \Elk proof and the detailed proof, respectively.
The reason is that the proofs get shorter with simpler axioms and inferences. 

The diversity of the answers confirms our assumption that the best method for 
generating proofs is often subjective, which justifies having all the
different proof generation methods in \PLUGINNAME.
Some participants commented on the presentation and
interaction with the proofs in \Protege (\eg the possibilities to expand the entire proof directly and not one 
inference at a time, to abbreviate names, and to choose between 
alternative forms of the same expression, etc.). Some of these comments have already been 
addressed in our tool \Evonne.

\setlength{\fboxrule}{1pt}
\setlength{\fboxsep}{0pt}

\section{Evonne}

The proof visualization and interaction techniques in \PLUGINNAME are restricted
by the capabilities of \Protege's explanation service. More advanced proof
exploration is offered by \Evee's big sister \Evonne~\cite{IJCAR22, 
DBLP:conf/dlog/AlrabbaaBDFK20,DBLP:conf/semweb/FlemischLAD20}, a web application 
that can be tried online and 
installed locally using Docker.\footnote{\url{https://www.docker.com/}}
\Evonne visualizes proofs with various interaction components and layouts to support the 
exploration of large proofs. The proof visualization is linked to a second view showing the context 
of the proof in the ontology by visualizing a subset (module) of the ontology that is relevant to 
the proof. We only give an overview of the proof visualisation here.

When starting \Evonne, users need to create a new project and associate an OWL
ontology file. They can then select an entailed atomic concept inclusion to be explained, and 
choose between the different proof generation methods described in 
Section~\ref{Sec:proof-types}. In addition, users can choose to provide a set of terms that they 
are familiar with (a signature) -- axioms using only those terms are then assumed to be known to 
the user, and consequently are not explained in the proof view (see~\cite{IJCAR22} for 
more details on this).

\begin{figure}
    \centering
    \fbox{
        \ifextendedVersion
        \includegraphics[width=.99\textwidth]{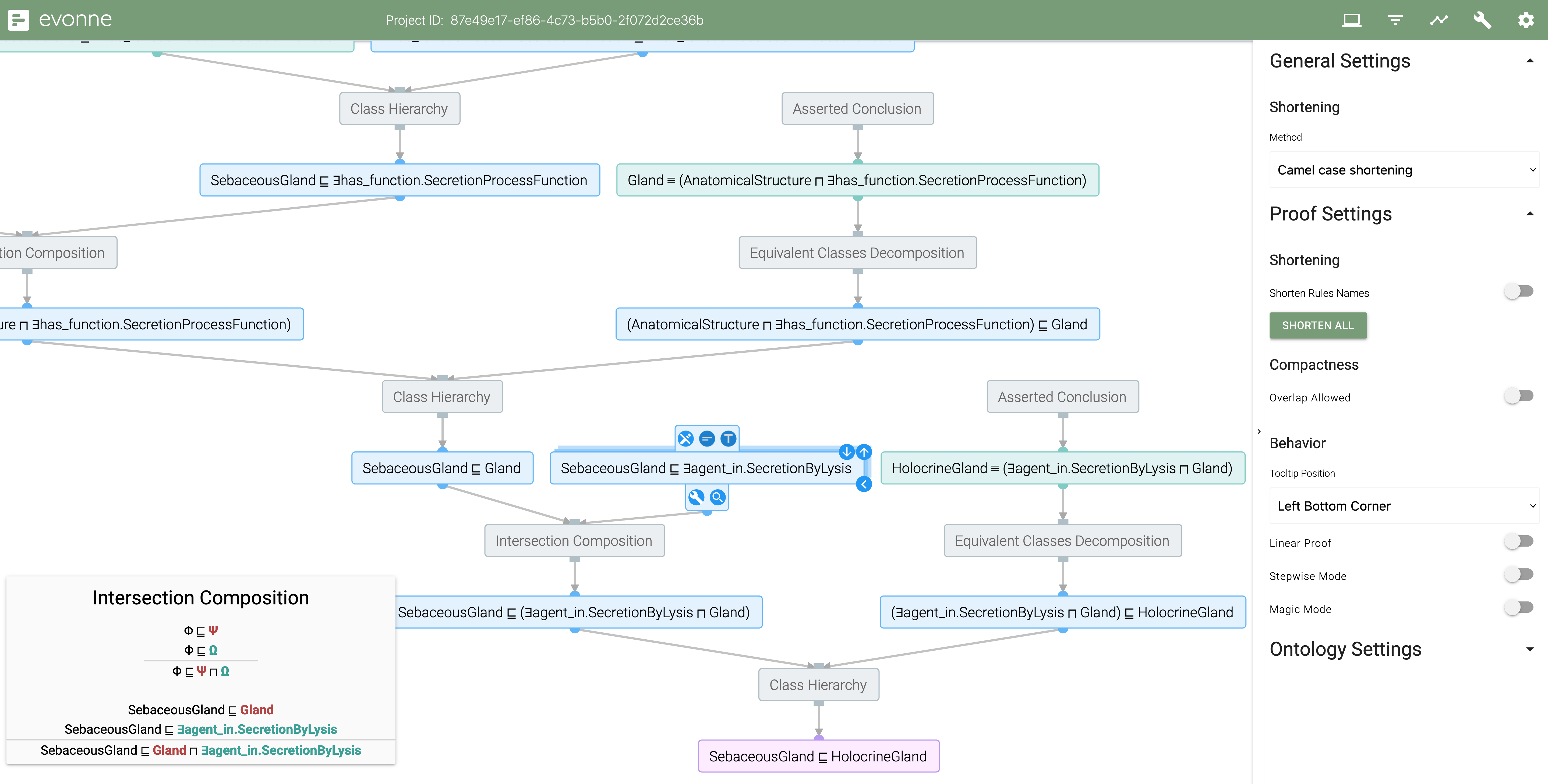}
        \else
        \includegraphics[width=.95\textwidth]{./screenshots/Evonne-Overview.png}
        \fi
    }
    \caption{\Evonne: Overview of the proof visualization component}
    \label{Fig:evonne}
\end{figure}

Proofs are shown as graphs with two kinds of vertices: colored vertices for axioms, gray
ones for inference steps. By default, proofs are illustrated using a \emph{tree layout}.
However, it is also possible to present proofs in a \emph{vertical layout}, placing axioms linearly 
below each other, with inferences represented through edges on the side (without the inference 
vertices). 
It is possible to automatically re-order vertices to minimize the distance between conclusion and 
premises in each step.
Another layout option is a  \emph{bidirectional layout}, which is a tree layout where, initially, the 
entire proof is collapsed into a \emph{magic vertex} that links the conclusion directly to its 
justification, and from which individual inference steps can be pulled out and pushed back from 
both directions. 
In all views, each vertex is equipped with multiple functionalities for exploring a proof.
For proofs generated using \Elk, clicking on an inference vertex shows the
inference rule used, and the particular inference with relevant sub-elements highlighted in 
different colors (see Figure~\ref{Fig:evonne}).
From each axiom vertex, users can alter the visual structure of the proof by
hiding
branches, revealing the previous inference step from the current node or showing the 
entire proof. Each of the different proof layouts has specific controls in this regard~-~for 
example, the vertical layout does not show inference rules as nodes and instead allows 
inspection on demand. Additionally, trays of buttons above and below nodes can be used %
to
manipulate the representation of axioms (by shortening names or using
natural language), and to
highlight justifications and diagnoses in the ontology
view~\cite{DBLP:conf/dlog/AlrabbaaBDFK20}.

\section{Conclusion}

We have presented a collection of methods for generating DL proofs that can be explored in various ways.
Future work includes adding support for specifying a \enquote{known} signature to \PLUGINNAME, the development of proof generation methods for even more expressive logics (\eg using consequence-based reasoners such as Sequoia~\cite{DBLP:conf/kr/BateMGSH16}), and visualizing approaches for explaining \emph{missing} entailments such as counter-interpretations~\cite{DBLP:conf/ki/AlrabbaaHT21} and abduction~\cite{DBLP:conf/kr/HaifaniKT21}.

\begin{acknowledgments}
  This work was supported by DFG grant 389792660 as part of TRR 248 – CPEC, and the DFG 
  Research Training Group QuantLA,  GRK 1763.
\end{acknowledgments}

\ifextendedVersion
\pagebreak
\appendix
\section{Appendix}

\subsection{Proofs Used in the User Study}

\begin{figure}[th]
    \begin{minipage}{.49\textwidth}
        \fbox{\includegraphics[width=\textwidth]{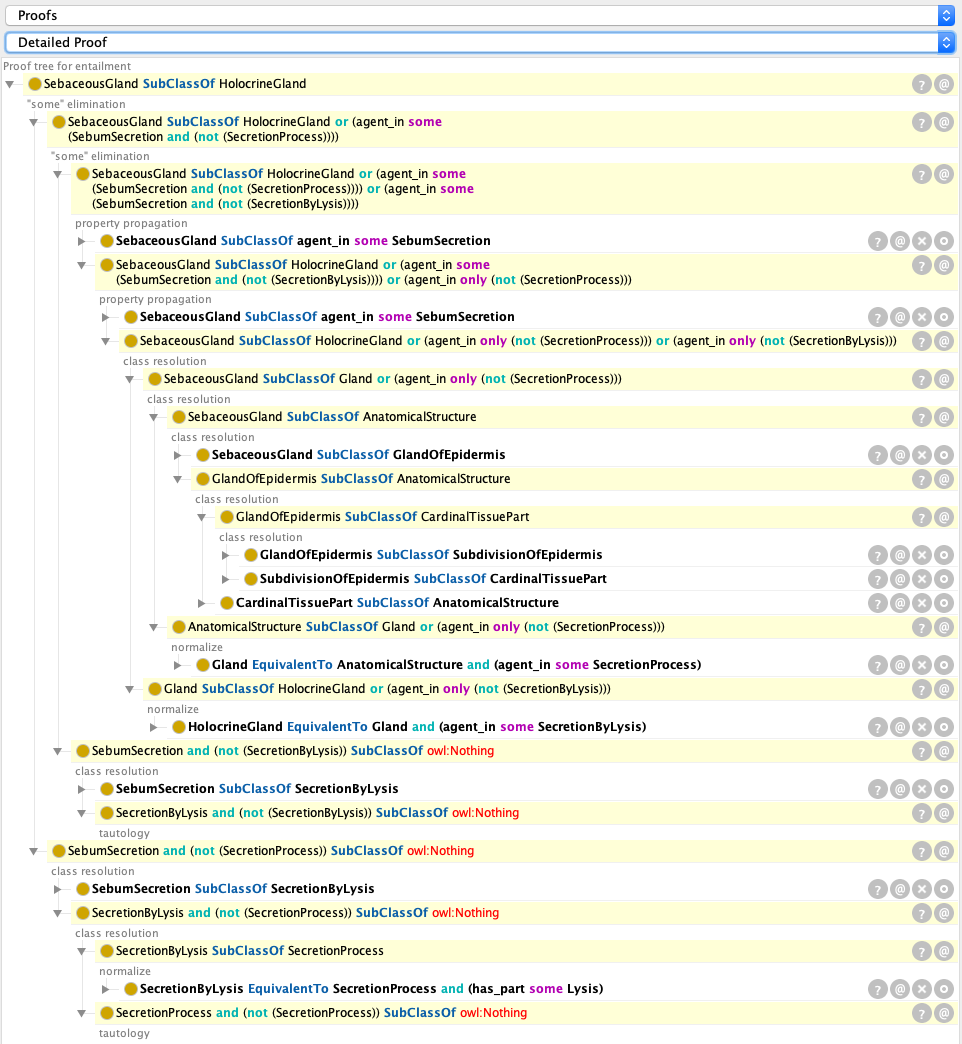}}
    \end{minipage}%
    \hfil%
    \begin{minipage}{.49\linewidth}
        \fbox{\includegraphics[width=\textwidth]{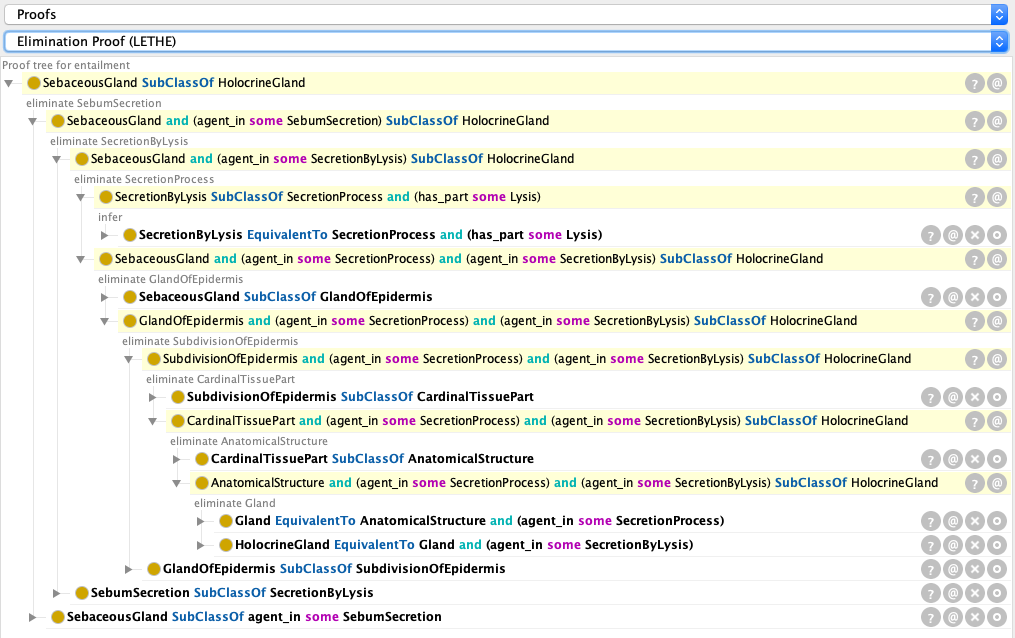}}
\end{minipage}
\caption {Task 1 ($\textsf{SebaceousGland} \sqsubseteq \textsf{HolocrineGland}$): Detailed 
Proof vs. Elimination Proof (\Lethe)}
\end{figure}

\begin{figure}[th]
    \begin{minipage}{.49\textwidth}
        \fbox{\includegraphics[width=\textwidth]{./userStudy/proofs/t1-EPL}}
    \end{minipage}%
    \hfil%
    \begin{minipage}{.49\linewidth}
        \fbox{\includegraphics[width=\textwidth]{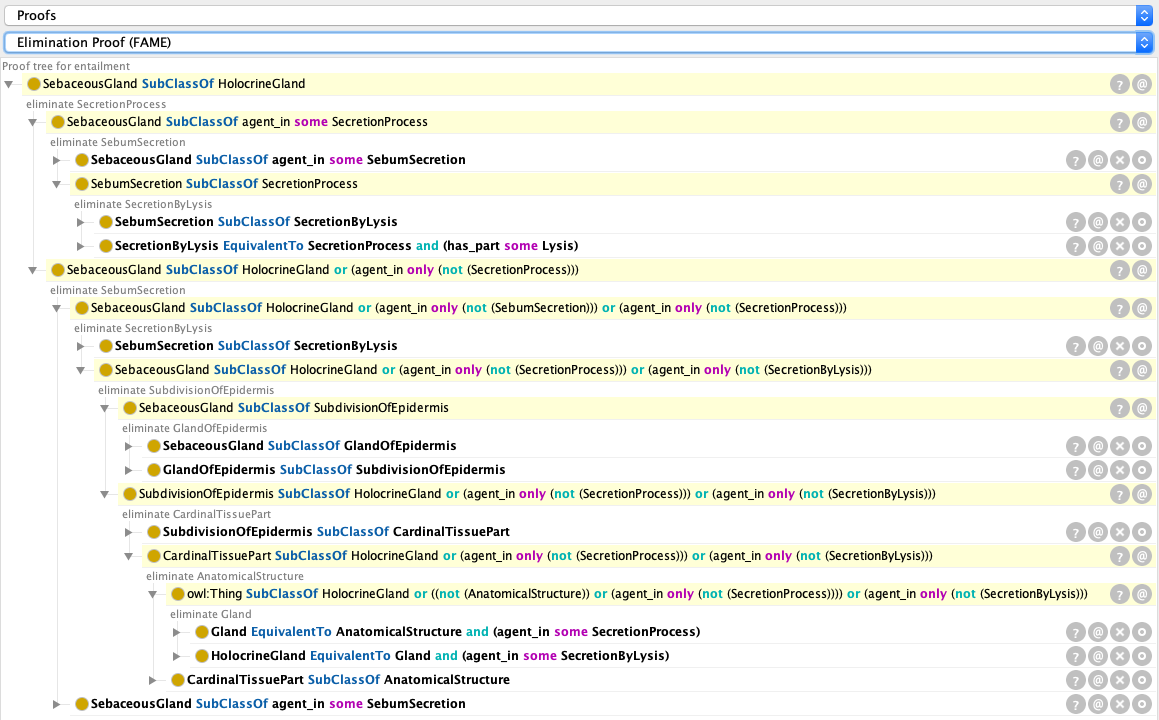}}
    \end{minipage}
    \caption {Task 2 ($\textsf{SebaceousGland} \sqsubseteq \textsf{HolocrineGland}$): 
    Elimination Proof (\Lethe) vs. Elimination Proof (\Fame)}
\end{figure}

\begin{figure}[th]
    \begin{minipage}{.49\textwidth}
        \fbox{\includegraphics[width=\textwidth]{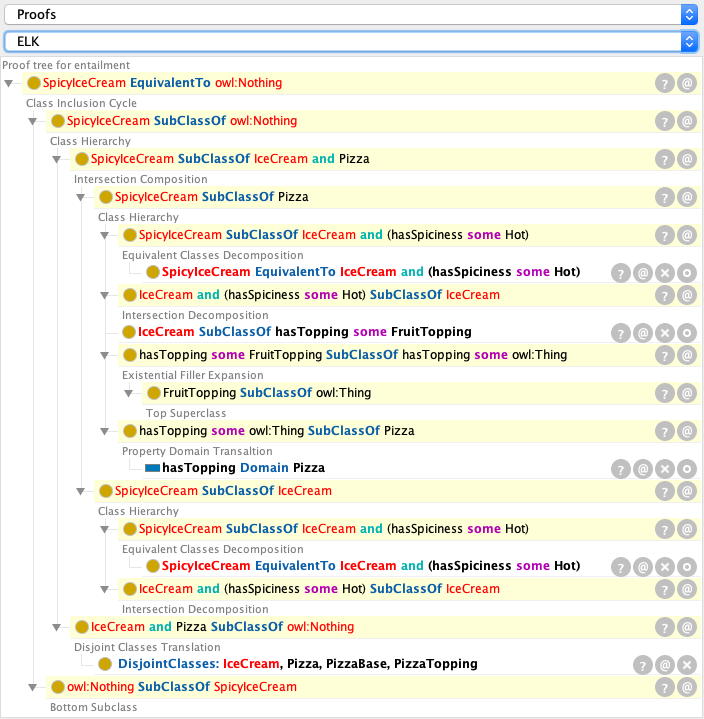}}
    \end{minipage}%
    \hfil%
    \begin{minipage}{.49\linewidth}
        \fbox{\includegraphics[width=\textwidth]{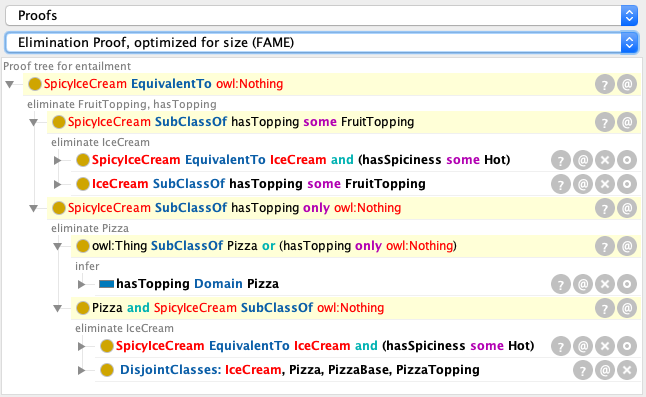}}
    \end{minipage}
    \caption {Task 3 ($\textsf{SpicyIceCream}\equiv \bot$): \Elk Proof vs. Elimination Proof, 
    Optimized For Size (\Fame)}
\end{figure}

\begin{figure}[th]
    \begin{minipage}{.49\textwidth}
        \fbox{\includegraphics[width=\textwidth]{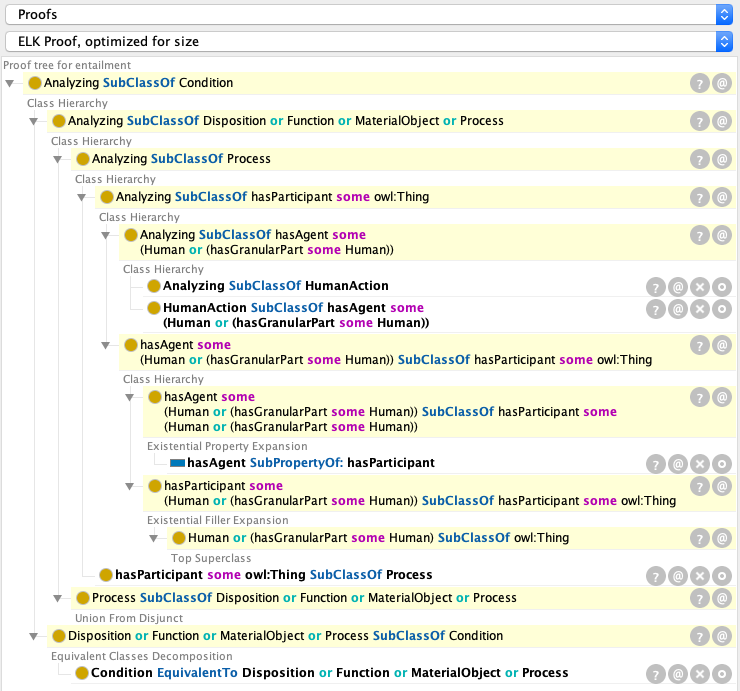}}
    \end{minipage}%
    \hfil%
    \begin{minipage}{.49\linewidth}
        \fbox{\includegraphics[width=\textwidth]{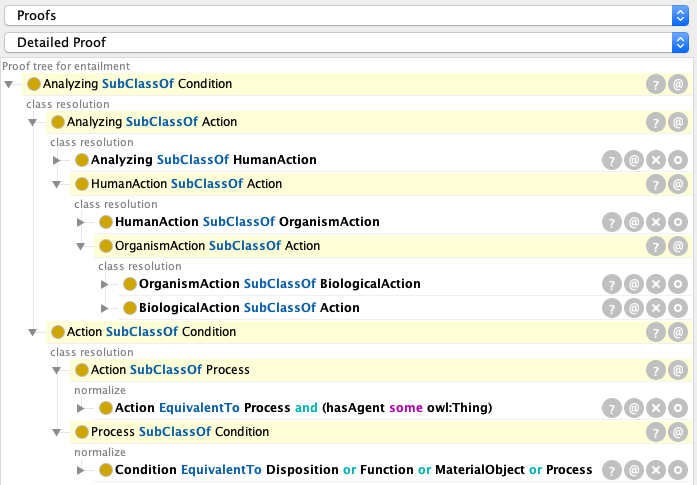}}
    \end{minipage}
    \caption {Task 5 ($\textsf{Analyzing} \sqsubseteq \textsf{Condition}$): \Elk Proof, 
    Optimized For Size vs. Detailed Proof}
\end{figure}

\begin{figure}[th]
    \begin{minipage}{.49\textwidth}
        \fbox{\includegraphics[width=\textwidth]{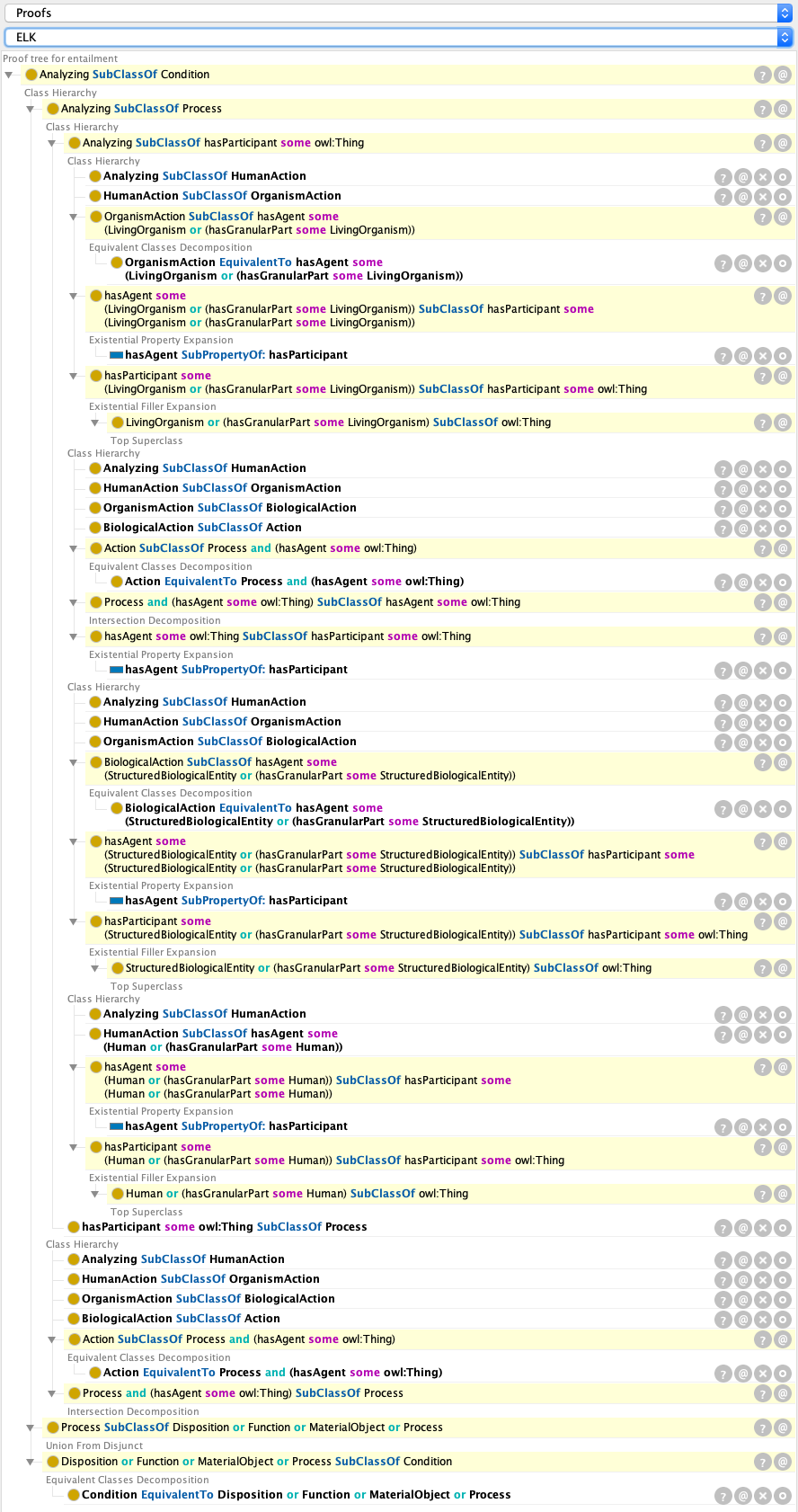}}
    \end{minipage}%
    \hfil%
    \begin{minipage}{.49\linewidth}
        \fbox{\includegraphics[width=\textwidth]{./userStudy/proofs/t4-ElkO}}
    \end{minipage}
    \caption {Task 4 ($\textsf{Analyzing} \sqsubseteq \textsf{Condition}$): \Elk Proof vs. \Elk 
    Proof, Optimized For Size}
\end{figure}

\fi

\end{document}